\documentclass[english,aps]{revtex4}
\usepackage[T1]{fontenc}
\usepackage[latin9]{inputenc}
\usepackage{array}
\usepackage{textcomp}
\usepackage{bm}
\usepackage{multirow}
\usepackage{amsmath}
\usepackage{amssymb}
\usepackage{graphicx}
\usepackage{esint}

\makeatletter

\providecommand{\tabularnewline}{\\}

\@ifundefined{textcolor}{}
{%
 \definecolor{BLACK}{gray}{0}
 \definecolor{WHITE}{gray}{1}
 \definecolor{RED}{rgb}{1,0,0}
 \definecolor{GREEN}{rgb}{0,1,0}
 \definecolor{BLUE}{rgb}{0,0,1}
 \definecolor{CYAN}{cmyk}{1,0,0,0}
 \definecolor{MAGENTA}{cmyk}{0,1,0,0}
 \definecolor{YELLOW}{cmyk}{0,0,1,0}
}

\makeatother

\usepackage{babel}
\begin{document}

\title{Enhancement of Vibronic and Ground-State Vibrational Coherences in
2D Spectra of Photosynthetic Complexes}

\author{Aur\'{e}lia Chenu$^{1}$, Niklas Christensson$^{2}$, Harald F.
Kauffmann$^{2}$, and Tom\'{a}\v{s} Man\v{c}al$^{1}$}

\address{$^{1}$Faculty of Mathematics and Physics, Charles University, Ke
Karlovu 5, 121 16 Prague 2, Czech Republic}

\address{$^{2}$Faculty of Physics, University of Vienna, Strudlhofgasse 4,
1090 Vienna, Austria}
\begin{abstract}
A vibronic-exciton model is applied to investigate the mechanism of
enhancement of coherent oscillations due to mixing of electronic and
nuclear degrees of freedom recently proposed as the origin of the
long-lived oscillations in the 2D spectra of the Fenna-Mattews-Olson
(FMO) pigment protein complex {[}Christensson \emph{et al.} J. Phys.
Chem. B {\bf 116} (2012) 7449{]}. Here we reduce the problem to a
model bacteriochlorophyll (BChl) dimer to elucidate the role of resonance
coupling, site energies, vibrational frequency and energy disorder
in the enhancement of vibronic-exciton coherences as well as ground-state
vibrational wave-packets, and to identify regimes where this enhancement
is significant. For an heterodimer representing the two relatively
strongly coupled BChls 3 and 4 of the FMO complex, we find that the
initial amplitude of the vibronic-exciton and vibrational coherences
are enhanced by up to 15 and 5 times, respectively, compared to the
vibrational coherences in the isolated monomer. This maximum initial
amplitude enhancement occurs when there is a resonance between the
electronic energy gap and the frequency of the vibrational mode. The
bandwidth of this enhancement is found to be about 100 cm$^{-1}$
for both mechanisms. The excitonic mixing of electronic and vibrational
degrees of freedom leads to additional dephasing relative to the ground-state
vibrational coherences. Here we evaluate the dephasing dynamics by
solving the quantum master equation in Markovian approximation and
observe a strong dependence of the life-time enhancement on the mode
frequency. It is found that long-lived vibronic-exciton coherences
are only generated when the frequency of the mode is in the vicinity
of the electronic resonance. Although the vibronic-exciton coherences
exhibit a larger initial amplitude compared to the ground-state vibrational
coherences, we conclude that both type of coherences have a similar
magnitude at longer population time for the present model. The ability
to distinguish between vibronic-exciton and ground-state vibrational
coherences in the general case of a coupled multi-chromophore system
is discussed. 
\end{abstract}
\maketitle

\section{Introduction}

The observation of long-lived oscillations in the two-dimensional
spectra of Fenna-Mattews-Olson (FMO) pigment-protein complex \cite{Engel2007a}
has renewed the general interest in quantum effects in light-harvesting
and other biological systems. Although similar observations were reported
previously \cite{Prokhorenko2002a,Savikhin1997a}, the interpretation
of the oscillations as electronic coherence and the suggestion that
such dynamical coherence may play a crucial role in achieving the
high efficiency of electronic energy transfer in photosynthesis \cite{Engel2007a}
have ushered a wave of research on the role of coherence in excitation
energy transfer processes. 

Numerous theoretical articles have been devoted to understanding and
explaining the long-lived oscillations in FMO. They have shown that
coherent oscillations last for several hundreds of femtoseconds, but
under standard assumptions about the properties of the protein environment,
none of these works have been able to account for the over picosecond
dephasing times of the oscillations observed experimentally. The only
proposed mechanism which predicts electronic coherence with life times
>1 ps is the correlation between environmental fluctuations on different
pigments \cite{Lee2007a,Rebentrost2009a,Ishizaki2010a,Hayes2010a,Abramavicius2011a,Caycedo-Soler2012a,Caram2012a}.
Experimental confirmation of correlated distributions of pigment energies
have been claimed for FMO \cite{Fidler2012a}, but neither dynamic
nor static correlation of the pigment energies have been found in
molecular dynamics (MD) simulations of the FMO protein environment
\cite{Olbrich2011a,Shim2012a}. Furthermore, dynamic correlation,
if present, would lead to an effective decrease of the system bath
coupling strength, and to a corresponding slowdown of the energy transfer
rates. Such a decrease of the system-bath coupling strength is at
odds with works which find an optimal strength of the system bath
interaction for transport function within the parameter range used
by standard theories \cite{Wu2010a,Shim2012a}. Also, the energy transfer
rates obtained by standard theory are in a good agreement with experimental
data \cite{Cho2005a}, and the standard \textquotedbl{}funnel picture\textquotedbl{}
of energy transfer \cite{BlankenshipBook} seems therefore to be well
supported.

The assumption that the experimentally observed coherences would be
relevant to the biological function of the antenna complexes has also
been criticized. Such a proposal requires the excitation with coherent
superpositions of states created by lasers in the laboratory to be,
in some way, equivalent to the excitations under\emph{ in vivo} conditions
\cite{Cheng2009a}, i.e. by direct sunlight or via transfer from another
antenna system. However, serious objections to this view have been
raised in the literature, arguing that direct excitation by light
from the sun (thermal light) does not lead to such coherent excitation
\cite{Jiang1991a,Mancal2010a,Brumer2011a,Hoki2011a,Pachon2012b}.
A recent experiment showed that even under coherent excitation of
the chlorosome antenna, i.e. with femtosecond laser pulses, no coherent
oscillations could be observed \cite{Dostal2012a}. This implies that
any coherence induced by the excitation would decay quickly, and that
the energy is transferred to the FMO complex in an incoherent fashion
independently of the excitation conditions.

The above discussion suggests that the mechanism explaining the oscillations
does not need to have a strong impact on the energy transfer dynamics
or be relevant to the energy transfer efficiency. The problem should
therefore be studied from the point of view of the experimental signals
rather than the (electronic) excited-state dynamics. In other words,
the origin of the oscillating experimental signal has to be understood
before the signal can be translated into claims about the energy transfer
dynamics. Recently, Christensson \textit{et al.} \cite{Christensson2012a}
proposed that the excitonic interaction between electronic and vibrational
states in FMO serves to create vibronic states (excitonically mixed
electronic and vibrational states). Such states have a considerable
vibrational character, and at the same time have an enhanced transition
dipole moment due to intensity borrowing from the strong electronic
transitions. It was also shown that coherent excitation of the vibronic
states produces oscillations in the non-linear signal that exhibit
picosecond dephasing times. Moreover, the concept of vibronic excitons
provides a plausible explanation for the observation of correlated
distribution of site energies \cite{Fidler2012a}. The vibronic-exciton
states involved in the long-lived coherences are to a large extent
composed of different vibrational states on the same pigment. This
automatically leads to a correlation in the fluctuations of the involved
transitions even for a random distribution of pigment energies. Finally,
the vibronic-exciton model has been shown to provide, compared to
the purely excitonic model, more realistic relaxation rates, as verified
against experimental measurements in two cyanobacterial light-harvesting
proteins \cite{Womick2011a}. 

In a recent work, Tiwari \textit{et al.} \cite{Tiwari2012a} used
a similar model to show that the mixing of electronic and vibrational
DOF leads to an enhancement of the excitation of vibrational coherences
in the electronic ground state as well, and it was also argued that
this effect can explain the long-lived oscillations in FMO. In order
to distinguish between the coherences observed in the excited-state
manifold and the ones originating in the ground-state manifold, we
will assign the term \textsl{vibrational} coherence strictly to the
latter ground-state contribution. There, the states are of pure vibrational
origin although the ability to excite them is enabled by the mixing
of vibrational and electronic states in the excited-state manifold.
When referring to the excited-state coherences with a strong vibrational
involvement, we will use the terms \textsl{vibronic}, or more specifically
\textsl{vibronic-exciton} coherences. 

In this paper, we turn to a model system (an FMO-inspired molecular
dimer) in order to systematically investigate how the mixing of vibrational
and electronic DOF leads to long-lived oscillatory signal in non-linear
optical spectroscopy. The paper is organized as follows. In the next
section, we explain the mechanism of coherence amplitude enhancement
on a toy model. In Section \ref{sec:Dimer-Hamiltonian:-Full} we introduce
the model dimer system with the associated notation, and we define
a measure of the vibrational character of a coherence in Section \ref{sec:Measure-of-the}.
Section \ref{sec:2D-Non-Linear-Spectroscopy} introduces the third-order
non-linear signal resolved in the coherent 2D Fourier transformed
spectroscopy and its relation to the dynamics of molecular systems,
in particular to its coherent dynamics. After determining the basic
properties of the enhancement for vibronic (excitonically mixed electronic
and vibrational) and the ground-state vibrational coherences in Section
\ref{sec:FMO-Inspired-Heterodimer}, we discuss the results in context
of FMO and the low frequency vibrational spectrum of BChl-a in Section
\ref{sec:Dimer-vs-FMO}.

\section{Enhancement of Coherence Amplitude by Transition Dipole Moment Borrowing:
A Toy Model\label{sec:Enhancement-of-Coherence}}

\begin{figure}
\includegraphics[width=1\columnwidth]{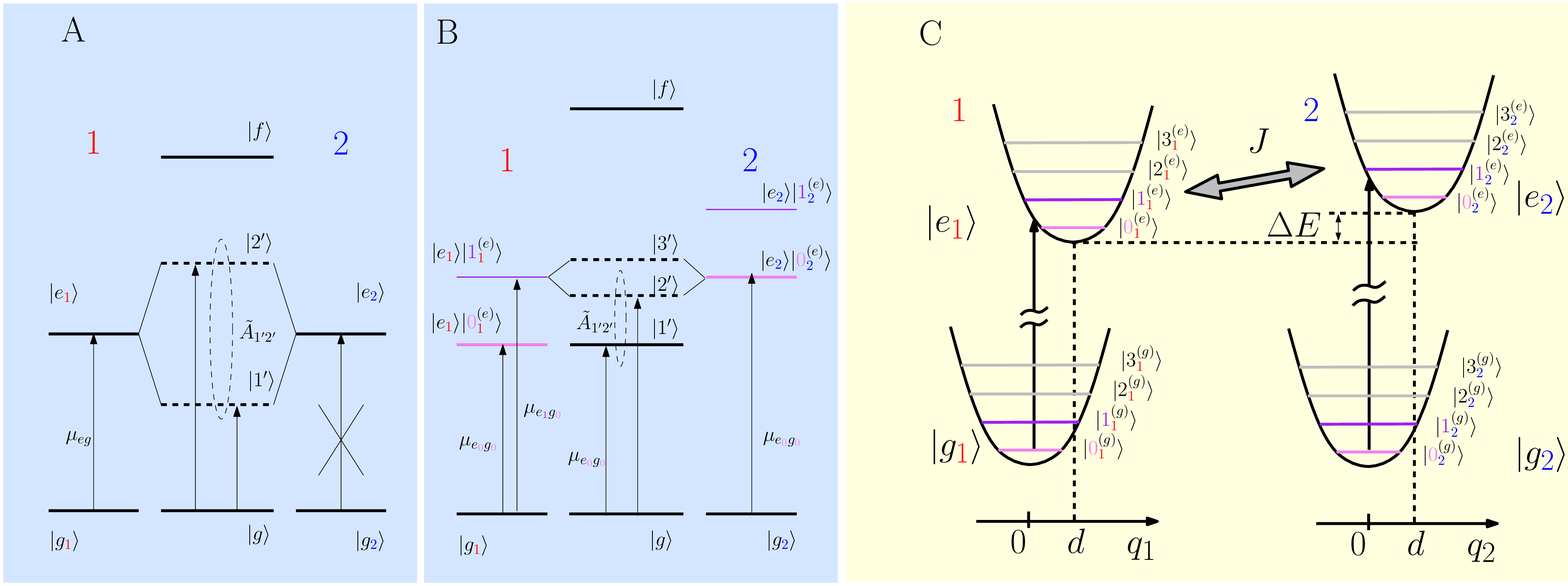}

\caption{\label{fig:Illustration}Illustration of the effect of transition
dipole moment borrowing and excitonic mixing of exited states in a
dimer without considering vibrational states (A) and considering a
simplified model of vibrational presence (B). In the case (B), a heterodimer
is considered with an approximate resonance between the one-phonon
state of monomer 1 and the zero-phonon state of monomer 2. Panel C:
Dimer model with a single vibrational mode per monomer. The monomers
1 and 2 can be in their respective ground states $|g\rangle$ or excited
states $|e\rangle$. In each electronic state, the monomers can occupy
any of the vibrational levels corresponding to their respective vibrational
modes ($q_{1}$ or $q_{2}$).}
\end{figure}

To illustrate the mechanism of enhancement of the coherence amplitude
that is suggested in Ref. \cite{Christensson2012a} and studied in
detail in this paper, let us first consider qualitatively a rather
trivial example. Let us imagine a system of two molecules, one of
which has a forbidden transition to the excited state. In the absence
of excitonic coupling between the excited states of the two molecules,
it is not possible to excite a linear combination of the collective
excited (eigen)states of the dimer (i.e. the coherence observed by
non-linear spectroscopy), because one of the transitions is forbidden.
If we now switch on the interaction between the monomers, it becomes
possible to excite the coherence between the new eigenstates of the
system (see Fig. \ref{fig:Illustration}A for illustration). We assume
the transition dipole moment in a form
\begin{equation}
\hat{\mu}=\mu_{eg}|1\rangle\langle0|+h.c.,
\end{equation}
where $|1\rangle=|e_{1}\rangle|g_{2}\rangle$ is the allowed collective
excited state of the dimer with the monomer 1 in its excited state
$|e_{1}\rangle$ and monomer 2 in its ground state $|g_{2}\rangle$
and $|0\rangle=|g_{1}\rangle|g_{2}\rangle$ is the collective ground
state. Expressed in the new eigenstate basis of vectors $|\alpha\rangle$,
the transition dipole moment reads
\begin{equation}
\hat{\mu}=\sum_{\alpha=1',2'}\left(\mu_{eg}\langle\alpha|1\rangle\right)|\alpha\rangle\langle0|+h.c..
\end{equation}
The new transition dipole moment to the excited eigenstates $|\alpha\rangle$
reads $\tilde{\mu}_{\alpha0}=\mu_{eg}\langle\alpha|1\rangle$. As
will become clear in the following sections, the amplitude of the
contribution of a coherence between two states $|\alpha\rangle$ and
$|\beta\rangle$ to the 2D spectrum is proportional to the factor
$\tilde{A}_{\alpha\beta}=|\tilde{\mu}_{\alpha0}|^{2}|\tilde{\mu}_{\beta0}|^{2}$.
In a dimer, we have only states $|1^{\prime}\rangle$ and $|2^{\prime}\rangle$,
i.e. only one coherence, and we obtain
\begin{equation}
\tilde{A}_{1^{\prime}2^{\prime}}=|\mu_{eg}|^{4}|\langle1^{\prime}|1\rangle|^{2}|\langle2^{\prime}|1\rangle|^{2}.
\end{equation}
Because $|\langle1^{\prime}|1\rangle|^{2}+|\langle2^{\prime}|1\rangle|^{2}=1$,
the maximum value of $\tilde{A}_{1^{\prime}2^{\prime}}$ is equal
to $\frac{1}{4}|\mu_{eg}|^{4}$ when the mixing is maximum, i.e. $|\langle1^{\prime}|1\rangle|^{2}=\frac{1}{2}.$ 

In a dimer, the coefficient $\langle1^{\prime}|1\rangle$ is given
by a sine or cosine of the mixing angle $\vartheta=\arctan\left(\frac{2J}{\Delta E}\right)$,
where $J$ is the resonance coupling and $\Delta E$ is the difference
between energies of the interacting states \cite{MayKuehnBook}. When
$J$ is small, the mixing can only be substantial at the resonance,
i.e. when $\Delta E\rightarrow0$. The observation of coherence in
2D spectroscopy depends crucially on the ability to excite the two
involved excited states simultaneously. In the above case, it is enabled
by the resonance between excited states. 

In Fig. \ref{fig:Illustration}B, we present another interesting case.
We extend the previous toy model with vibrational states at each monomer,
and assume both monomers to have allowed transitions to the excited
states. In Fig. \ref{fig:Illustration}B, only the first vibrationally
excited state in the electronically excited state is represented.
The one-phonon state of monomer 1 interacts weakly with the zero-phonon
electronically excited state of monomer 2. Because the coupling is
small, the interaction between the two zero-phonon states of the interacting
monomers can be neglected, and the borrowing effect occurs only for
the collective states $|2^{\prime}\rangle$ and $|3^{\prime}\rangle$
of Fig. \ref{fig:Illustration}B. If, in addition, we assume the transition
dipole moments to the one-phonon state to be small (i.e. the vibrational
mode has a low Huang-Rhys (HR) factor), we obtain the situation similar
to Fig. \ref{fig:Illustration}A. At resonance (i.e. at maximum mixing
$|\langle2^{\prime}|e_{2}\rangle|^{2}\approx\frac{1}{2}$), the coherence
amplitude $\tilde{A}_{1^{\prime}2^{\prime}}$ in 2D spectrum reads
\begin{equation}
\tilde{A}_{1^{\prime}2^{\prime}}\approx|\mu_{e_{0}g_{0}}|^{2}|\mu{}_{e_{0}g_{0}}|^{2}\frac{1}{2}.
\end{equation}
This is to be compared with the amplitude of the purely vibrational
coherence on a non-interacting monomer which is $\tilde{A}_{vib}=|\mu_{e_{0}g_{0}}|^{2}|\mu_{e_{1}g_{0}}|^{2}$
and which is small due the small HR factor. The excitonic interaction
thus enhances a coherence which has partially both the vibrational
and electronic characters.

The situation studied later in this paper is similar to the model
illustrated in Fig. \ref{fig:Illustration}B. We will study coherences
involving the zero- and one-phonon states of the same vibrational
mode on a given monomer. As demonstrated, the transition from the
electronic and vibrational ground state to the one-phonon state can
be enhanced by excitonic coupling with the allowed states of other
monomers. This was found in Ref. \cite{Christensson2012a} to be the
case for FMO, where the enhanced coherences where found to be of a
prevailingly vibrational character, thus exhibiting significantly
prolonged life time over the purely electronic coherences. 

The excited-state mixing between the vibrational states of one monomer
and the electronic (zero-phonon) state of the other monomers can,
however, enhance also the ground-state bleaching signal \cite{Tiwari2012a},
which was not included in our toy model. The life-time of the coherence
has not been discussed within the toy model either, because it is
not relevant to the illustration of the enhancement mechanism. However,
it will be of crucial importance throughout the rest of the paper.

\section{Dimer Hamiltonian: Full Formulation\label{sec:Dimer-Hamiltonian:-Full}}

In this paper, we concentrate on the relatively strongly coupled dimer
of BChls 3 and 4 of the FMO complex \cite{Adolphs2006a}, which we
will later refer to as monomer 1 and monomer 2, respectively. Needless
to say, the interaction of the other BChls in the complex with this
dimer is decisive for its function as a molecular wire, and it influences
the properties of the oscillations observed in 2D spectra as well.
In Ref. \cite{Christensson2012a}, it was demonstrated that the effect
of long-lived coherence could plausibly originate from the interplay
between the vibrational modes local to BChl 3 and the resonance interaction
between BChls 3 and 4. By reducing FMO to this dimer, we expect to
be able to isolate the main quantitative contributions to the observed
effect, and to simplify the treatment such that the main conditions
for the mechanism to take place can be identified. The extension of
the results to the entire FMO complex is discussed in Section \ref{sec:Dimer-vs-FMO}. 

The studied molecular system will be represented by a dimer Hamiltonian,
explicitly treating a single intramolecular vibration (with frequency
$\omega_{0}$) on each of the monomers. Unlike in Ref. \cite{Christensson2012a},
we will first consider the formulation with an arbitrary number of
vibrational-excited levels, which we refer to as a full formulation
(full with respect to the number of vibrational levels). For practical
calculations, we will reduce the number of vibrational levels to a
single mode on each monomer. This is justified by the low Huang-Rhys
factors used in this work. The dimer Hamiltonian in the electronic
site basis is defined as
\begin{equation}
\hat{H}_{S}=\hat{H}^{(1)}\otimes\mathbb{I}^{(2)}+\mathbb{I}^{(1)}\otimes\hat{H}^{(2)}+J_{12}\left(|e_{1}\rangle|g_{2}\rangle\langle g_{1}|\langle e_{2}|+h.c.\right),\label{eq:Hamiltonian}
\end{equation}
where $\mathbb{I}^{(n)}$ represents the identity operator on the
Hilbert space of monomer $n$, and

\begin{equation}
\hat{H}^{(n)}=\left[\frac{\hbar\omega_{0}}{2}\left(\hat{p}_{n}^{2}+\hat{q}_{n}^{2}\right)\right]|g_{n}\rangle\langle g_{n}|+\left[\frac{\hbar\omega_{0}}{2}\left(\hat{p}_{n}^{2}+(\hat{q}_{n}-d)^{2}\right)\right]|e_{n}\rangle\langle e_{n}|,\label{eq:Hamiltonian_molecule}
\end{equation}
is the Hamiltonian of monomer $n$. Here $\hat{p}_{n}$ and $\hat{q}_{n}$
are the dimensionless momenta and coordinate operators of the vibrational
mode on site $n$, respectively, $d$ is the coordinate displacement,
and $J_{nm}$ denotes the inter-site coupling. The corresponding level
model including the vibrational states and the associated notation
is presented in Fig. \ref{fig:Illustration}C. 

This system interacts with a bath of protein degrees of freedom (DOF)
which is modeled as an infinite number of harmonic oscillators characterized
by some continuous spectral density, e.g. $J(\omega)=\sum_{a}g_{a}^{2}\delta(\omega-\omega_{a})$.
The Hamiltonian representing the coupling between the bath modes characterized
by the Hamiltonian
\begin{equation}
\hat{H}_{B}=\sum_{a\neq0}\frac{\hbar\omega_{a}}{2}\left(\hat{P}_{a}^{2}+\hat{Q}_{a}^{2}\right),
\end{equation}
and the dimer reads as
\begin{equation}
\hat{H}_{I}=\sum_{n,a\neq0}\kappa_{n}^{(a)}\hat{Q}_{a}|e_{n}\rangle\langle e_{n}|+\sum_{n,a}k_{n}^{(a)}\hat{Q}_{a}\hat{q}_{n}.\label{eq:IntHam}
\end{equation}
The first term in Eq. (\ref{eq:IntHam}) corresponds to the linear
interaction between the bath and the electronic DOF and can be expressed
through the spectral density as $\kappa_{n}^{(a)}=\hbar\omega_{a}g_{a}^{(n)}$
\cite{Adolphs2006a}, while the second term in Eq. (\ref{eq:IntHam})
describes a bi-linear interaction between the intramolecular mode
and the bath. Following the work of Refs. \cite{Christensson2012a}
and \cite{Tiwari2012a}, we will consider this interaction to be weak
and set $k_{n}^{(a)}=0$ for all $n$, accordingly.

Let us now define the notation used throughout this paper to describe
the dimer state. As in the standard electronic molecular exciton theory,
the excited states of the dimer are formed out of the excited states
of the monomers. We denote the electronic excited states by a diad
$(nm)$ where $n$ ($m$) stands for the state ($g$ or $e$) of monomer
1 (2) in correspondence with Fig. \ref{fig:Illustration}C. The Hilbert
space in which the molecular Hamiltonian will be represented can be
fully described by states $|n_{\nu}m_{\nu^{\prime}}\rangle,$ where
the first (second) letter describes the first (second) monomer's electronic
state, and the index $\nu,\,\nu^{\prime}$ denote their respective
vibrational quantum level, so that e.g.
\begin{equation}
|g_{\nu}g_{\nu^{\prime}}\rangle=|g_{1}\rangle|\nu_{1}^{(g)}\rangle|g_{2}\rangle|\nu_{2}^{\prime(g)}\rangle,
\end{equation}
\begin{equation}
|g_{\nu}e_{\nu^{\prime}}\rangle=|g_{1}\rangle|\nu_{1}^{(g)}\rangle|e_{2}\rangle|\nu_{2}^{\prime(e)}\rangle.
\end{equation}
The eigenstates of the Hamiltonian will be written as linear combinations
of the states $|n_{\nu}m_{\nu^{\prime}}\rangle$, i.e.
\begin{equation}
|\alpha\rangle=\sum_{n\neq m}\sum_{\nu,\nu^{\prime}}c_{n_{\nu}m_{\nu^{\prime}}}^{\alpha}|n_{\nu}m_{\nu^{\prime}}\rangle.
\end{equation}
The expansion coefficients obtained by diagonalization of the Hamiltonian
\begin{equation}
c_{n_{\nu}m_{\nu^{\prime}}}^{\alpha}\equiv\langle\alpha|n_{\nu}m_{\nu^{\prime}}\rangle
\end{equation}
 will report on the participation of the particular local states in
the eigenstates. The eigenstates will be numbered with increasing
energy $\epsilon_{\alpha}$, $\alpha=1,2,\dots$.

\section{Measure of the vibrational character\label{sec:Measure-of-the}}

To measure the vibrational character of a coherence, we can use the
square of the expansion coefficients. We will be interested in the
coherence between states $|e_{0}g_{\nu^{\prime}}\rangle$ and $|e_{1}g_{\nu^{\prime}}\rangle$.
This coherence is a vibrational coherence located on monomer 1, and
we ignore the vibrational state of the second monomer. Because the
ground-state vibrational quanta of the second monomer cannot be excited
from the thermal ground state by a single interaction with the exciting
pulse in the optical domain, we can set $\nu^{\prime}=0$. We are
interested in identifying which coherences between the eigenstates
have the character of this local vibrational coherence, and we want
to define accordingly a measure $\chi_{\alpha\beta}$ of the \emph{local
vibrational character }of the coherence $\rho_{\alpha\beta}$. For
a given coherence to be vibrational on monomer 1, the state $|\alpha\rangle$
needs to have a character of the zero-phonon state on the monomer
1, i.e. it has to be predominantly composed of the state $|e_{0}g_{0}\rangle$,
and the state $|\beta\rangle$ has to have predominantly the character
of the state $|e_{1}g_{0}\rangle$ (or \emph{vice versa}). This composition
is measured by the square of the corresponding expansion coefficients
$c_{n_{\nu}m_{\nu^{\prime}}}^{\alpha}$ and $c_{n_{\nu}m_{\nu^{\prime}}}^{\beta}$,
which also define the probability of finding the system in the corresponding
local states, should we attempt such a measurement. The character
of the coherence corresponds to the conditional probability of finding
the state $|e_{0}g_{0}\rangle$ by measuring on state $|\alpha\rangle$,
and of simultaneously finding the state $|e_{1}g_{0}\rangle$ by measuring
on state $|\beta\rangle$, or \emph{vice versa}. This leads us naturally
to the definition 
\begin{equation}
\chi_{\alpha\beta}\equiv\chi_{\alpha\beta}^{(e_{0}g_{0},e_{1}g_{0})}=|c_{e_{0}g_{0}}^{\alpha}|^{2}|c_{e_{1}g_{0}}^{\beta}|^{2}+|c_{e_{1}g_{0}}^{\alpha}|^{2}|c_{e_{0}g_{0}}^{\beta}|^{2}.\label{eq:vibchar}
\end{equation}
This quantity has its maximum when the eigenstates are each entirely
composed of the zero- or one-phonon states of the excited molecule,
and thereby correspond to the site basis states ($\chi_{\alpha\beta}=1$
for $|\alpha\rangle=|e_{0}g_{0}\rangle$ and $|\beta\rangle=|e_{1}g_{0}\rangle,$
or \emph{vice versa}). When the states $|\alpha\rangle$ and $|\beta\rangle$
are equal mixtures of these local states, so that $|c_{e_{0}g_{0}}^{\alpha}|^{2}=|c_{e_{1}g_{0}}^{\alpha}|^{2}=|c_{e_{0}g_{0}}^{\beta}|^{2}=|c_{e_{1}g_{0}}^{\beta}|^{2}=\frac{1}{2}$,
then we have $\chi_{\alpha\beta}=\frac{1}{2}$. Therefore, coherences
$\rho_{\alpha\beta}$ with $\chi_{\alpha\beta}>\frac{1}{2}$ will
be considered to have \emph{prevailingly local vibrational character}.
We compared the local vibrational character defined by Eq. (\ref{eq:vibchar})
with the composition of the eigenstate vectors $|c_{n_{\nu}m_{\nu^{\prime}}}^{\alpha}|^{2}$
, and we verified that our conclusions remain the same regardless
of the quantity used for the definition of the coherence character.
Note that the character $\chi_{\alpha\beta}$ is a time-independent
quantity. It refers to eigenstates of the Hamiltonian and their representation
in the basis of states local to the chromophores. The coherences have
life times dependent of the interaction of the system with the bath,
but their local vibrational character $\chi_{\alpha\beta}$ only depends
on the system Hamiltonian and does not evolve in time.

\section{2D Non-Linear Spectroscopy and Liouville Pathways \label{sec:2D-Non-Linear-Spectroscopy}}

Non-linear spectroscopy has been widely used to study the dynamics
of excitonic energy transfer in light-harvesting systems, in particular
because it is sensitive to the time dependent redistribution of the
populations among excited states. Details about this class of measurement
techniques are outside the scope of this work and can be found elsewhere
in literature (see e.g. \cite{ParsonBook,DemtroederI2008a,DemtroederII2008a}).
Two-dimensional (2D) coherent spectroscopy, which is one of the recent
additions to this class of techniques, has enabled us to directly
observe the coherent components of the excited-state dynamics \cite{Jonas2003a,Pisliakov2006a}.
In 2D coherent electronic spectroscopy, the time dependent signal
(the detection time is usually denoted as $t_{3}$ here) is generated
from the interaction of the sample with three consecutive ultrafast
laser pulses with wave vectors $\bm{k}_{1}$, $\bm{k}_{2}$ and $\bm{k}_{3}$
separated by time intervals $t_{1}$ and $t_{2}$ \cite{Jonas2003a,Cho2008a}.
The signal is heterodyne detected, i.e. the generated field rather
than its intensity is measured, and it is spectrally resolved in frequency
$\omega_{3}$ ($t_{3}$ and $\omega_{3}$ are correspondingly related
by Fourier transform). The measured field is then numerically Fourier
transformed in time $t_{1}$ ($t_{1}\rightarrow\omega_{1}$) resulting
in a signal dependent on two frequencies $\omega_{3}$ and $\omega_{1}$
and one time delay $t_{2}$. 2D spectra are represented as the frequency-frequency
correlation plots at various time delays $t_{2}$. The 2D signal is
related to the third-order polarization of the molecular system generated
by the exciting laser pulses, and its theoretical description is most
conveniently based on the non-linear response function formalism in
the third-order of perturbation theory (see e.g. Ref. \cite{MukamelBook}).
In this section, we will briefly present the various signal components
that can be distinguished in a 2D spectrum, and we will give some
of their characteristics, such as positions in the 2D spectrum, oscillation
frequency, initial- and time-dependent amplitudes.

\begin{figure}
\includegraphics[width=0.9\columnwidth]{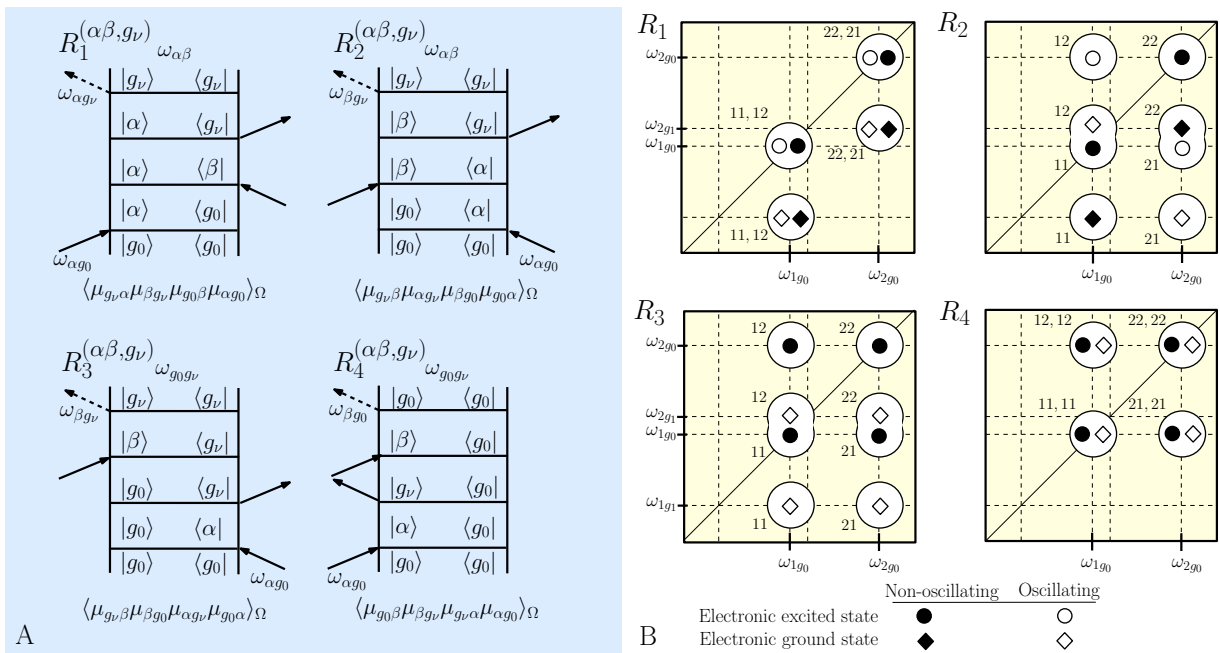}

\caption{\label{fig:Liouville-pathways}Liouville pathways and their respective
contribution to the 2D spectrum. (A) Double-sided Feynman diagram
describing the four Liouville pathways along with their dipole pre-factor,
the frequency of the involved coherence and the frequency of the exciting
and probing pulses. (B) Position of the signal in the 2D spectrum
and its characteristics originating from the different pathways. Solid
symbol denote non-oscillating contributions in $t_{2}$, open symbols
denote oscillatory contributions with frequencies $\omega_{\alpha\beta}$
(circle), $\omega_{g_{0}g_{\nu}}$ (diamond). Signals denoted by circle
symbols do not involve any vibrational level whereas those denoted
by diamonds do (for $\nu\geq1$). The diads in (B) represent the eigenstates
$|\alpha\rangle,$ $|\beta\rangle$ involved in each signal contribution. }
\end{figure}

Depending upon the time ordering of the light-matter interactions
in the third-order response functions, four different types of so-called
Liouville pathways (denoted $R_{i},\ i=1,\dots,4$) can be distinguished
in the generated signal. They are represented in Fig. \ref{fig:Liouville-pathways}A
by double-sided Feynman diagrams \cite{MukamelBook}. Pathways $R_{1}$
and $R_{4}$, i.e. with negative $t_{1}$, are of the so-called non-rephasing
character, while $R_{2}$ and $R_{3}$ are of the rephasing character.
Rephasing and non-rephasing signals can be measured separately. The
Liouville pathways can also be grouped together according to the electronic
band in which the coherences observed in 2D spectra originate. In
$R_{1}$ and $R_{2}$, the observed coherences stem from the electronically
excited state, while for $R_{3}$ and $R_{4}$, they come from the
ground state. Thus in both rephasing and non-rephasing pairs of Liouville
pathways, we observe one ground- and one excited-state coherence.
It is notoriously difficult to distinguish ground- and excited-state-originated
signal experimentally, unless one can find some secondary characteristics
(such as signal life time due to relaxation or frequency) which distinguishes
the ground- from the excited-state contributions.

Fig. \ref{fig:Liouville-pathways}B illustrates the position, origin
(excited or ground state) and oscillating frequency of the signal
generated by the different pathways in the 2D spectrum. The non-rephasing
pathways $R_{1}^{(\alpha\beta,g_{\nu})}$ and $R_{4}^{(\alpha\beta,g_{\nu})}$
result in peaks at the position $(\omega_{\alpha g_{0}},\,\omega_{\alpha g_{\nu}})$
and $(\omega_{\alpha g_{0}},\,\omega_{\beta g_{0}})$, respectively,
whereas the rephasing pathways $R_{2}^{(\alpha\beta,g_{\nu})}$ and
$R_{3}^{(\alpha\beta,g_{\nu})}$ generate peaks located in similar
positions, i.e. $(\omega_{\alpha g_{0}},\,\omega_{\beta g_{\nu}})$.
It is interesting to note the role of the index $\nu$ in the pathways.
For $\nu=0$, there are no ground-state coherences, while excited-state
coherences can appear. The frequency of oscillation of the peaks coming
from $R_{1}^{(\alpha\beta,g_{\nu})}$ and $R_{2}^{(\alpha\beta,g_{\nu})}$
is $\omega_{\alpha\beta}$, while it is given by the ground-state
vibrational frequency $\omega_{g_{\nu}g_{0}}$ for $R_{3}^{(\alpha\beta,g_{\nu})}$
and $R_{4}^{(\alpha\beta,g_{\nu})}$. For $\nu>0$, the amplitudes
of $R_{1}$ and $R_{4}$ pathways are equal, and the same holds for
$R_{2}$ and $R_{3}$ pathways. Thus, if we would ignore the effect
of the coherence life time, there would always be the same amplitude
of the ground- and excited-state oscillating contributions ($\nu>0$),
plus one contribution from the excited state ($\nu=0)$. 

Let us now discuss the (initial) amplitudes of oscillating features
in 2D spectra. The initial ($t_{2}=0$) amplitude of the third-order
response function is proportional to a transition dipole moment pre-factor
of the response function $R_{i}$ averaged over an isotropic distribution
of orientations of the molecules in the sample. The dipole pre-factors
$A^{(i)}$ for each Liouville pathway $R_{i}$ are defined in Appendix
\ref{sec:Third-Order-Response-Function}. For the non-rephasing response
function $R_{1}$, it reads in particular 
\begin{equation}
A_{\alpha\beta,g_{\nu}}^{(1)}=\langle(\bm{\mu}_{g_{\nu}\alpha}\cdot\bm{e}_{4})(\bm{\mu}_{\beta g_{\nu}}\cdot\bm{e}_{3})(\bm{\mu}_{g_{0}\beta}\cdot\bm{e}_{2})(\bm{\mu}_{\alpha g_{0}}\cdot\bm{e}_{1})\rangle_{\Omega}\label{eq:amplitude_R1}
\end{equation}
where $\bm{e}_{i}$ represents the unitary orientational vector of
the $i^{th}$ laser pulse, $\bm{\mu}_{\alpha g_{\nu}}$ is the eigenstate
transition dipole moment and $\left<\dots\right>_{\Omega}$ stands
for the averaging over an ensemble of randomly oriented complexes.
The disorder averaging can be accounted for by a constant orientation
factor $\Omega_{\alpha\beta}=\frac{1}{15}\left(2\cos^{2}{\theta_{\alpha\beta}}+1\right)$,
where $\theta_{\alpha\beta}$ is the angle between the transition
dipole moments of eigenstates $\alpha$ and $\beta$ \cite{Hochstrasser2001a}.
The amplitude pre-factor then reads
\begin{equation}
A_{\alpha\beta,g_{\nu}}^{(1)}=|\bm{\mu}_{g_{\nu}\alpha}||\bm{\mu}_{\beta g_{\nu}}||\bm{\mu}_{g_{0}\beta}||\bm{\mu}_{\alpha g_{0}}|\,\Omega_{\alpha\beta}.
\end{equation}
Here, $|\bm{\mu}|=\sqrt{\mu_{x}^{2}+\mu_{y}^{2}+\mu_{z}^{2}}$ is
the length of the transition dipole moment vector.

The time-dependent response function is composed of the dipole pre-factor
and the time-dependent part representing time evolution of the state
of the molecular systems during the time interval between the interactions
with light. The response functions $R_{i}$ for all pathways are presented
in detail in Appendix \ref{sec:Third-Order-Response-Function}. In
particular, the non-rephasing response function $R_{1}$ reads as
\begin{equation}
R_{\alpha\beta,g_{\nu}}^{(1)}(t_{3},t_{2},t_{1})=\left<A_{\alpha\beta,g_{\nu}}^{(1)}\, G_{\alpha g_{\nu}}(t_{3})\, G'_{\alpha\beta}(t_{2})\, G_{\alpha g_{0}}(t_{1})\right>_{\Delta}\label{eq:response_function_R1}
\end{equation}
where $G_{\alpha g_{\nu}}(t)$ and $G'_{\alpha\beta}(t)$ are the
evolution operators of the optical coherence $\rho_{\alpha g_{\nu}}$
and of the excited-state coherence $\rho_{\alpha\beta}$, respectively.
Here, $\langle\dots\rangle_{\Delta}$ denotes averaging over the energy
disorder. The details of the evolution operators are given in the
Appendix. The contribution of the corresponding coherence to the signal
in the 2D spectra is obtained after Fourier transformation of the
evolution propagators during the time intervals $t_{1}$ and $t_{3}$:

\begin{equation}
S_{\alpha\beta,g_{\nu}}^{(1)}(\omega_{3},t_{2},\omega_{1})=\left<A_{\alpha\beta,g_{\nu}}^{(1)}\:\tilde{G}_{\alpha g_{\nu}}(\omega_{3})\, G'_{\alpha\beta}(t_{2})\,\tilde{G}_{\alpha g_{0}}(\omega_{1})\right>_{\Delta}
\end{equation}
 with $\tilde{G}_{\alpha g_{\nu}}(\omega)=\int_{0}^{+\infty}dt\, e^{i(\omega-\omega_{\alpha g_{\nu}})t-\Gamma_{\alpha}t-\gamma_{\alpha\alpha}g(t)}$
and where $\omega_{1},$ $\omega_{3}$ are the excitation and probing
frequencies, respectively. The signal for the non-rephasing pathway
$R_{4},$ which involves the electronic ground-state coherence can
be obtained in the same way. Full expressions are given in the Appendix.

\section{FMO-Inspired Heterodimer in One Particle Approximation\label{sec:FMO-Inspired-Heterodimer}}

In order to quantify the enhancement of coherence amplitude by borrowing
of transition dipole moment, as illustrated in Section \ref{sec:Enhancement-of-Coherence},
we study a simplified molecular dimer model. In this section, we will
first study the initial amplitude of the dipole pre-factor for selected
coherences as a function of the molecule inter-site coupling $J$
and the energy gap $\Delta E$. We will include one particular nuclear
mode ($\omega_{0}=117$~cm$^{-1}$) at each of the two sites. Then,
we will investigate the dynamical behavior of the signal as a function
of the dimer energy gap and the vibrational mode frequency. Possible
consequences of our results for the case of FMO molecular aggregate
will be discussed in Section \ref{sec:Dimer-vs-FMO}. 

As mentioned above, the studied dimer is inspired by the FMO lowest
energy BChls 3 and 4 which are relatively strongly interacting (site
coupling $J_{0}=-53.5$~cm$^{-1}$) and form a heterodimer with an
energy gap $\Delta E_{0}=110$~cm$^{-1}$ (later referred to as the
reference energy gap) \cite{Adolphs2006a}. The directions of the
transition dipole moments were taken from the Protein Data Bank file
3ENI \cite{Tronrud2009a} (we used $\theta_{\textrm{Bchl 3-4}}$ =
107.2\textdegree{}). A Huang-Rhys factor of $S=0.05$ has been used.
In order to assess the influence of the nuclear modes on the dimer
spectra, we studied the dependence of the spectra on selected model
parameters for a single vibrational mode. We study two cases, with
frequencies $\omega_{0}=117$~cm$^{-1}$ and $185\ {\rm cm}^{-1}$.
For some parameters and spectral characteristics, we study the system
properties as a function of the mode frequency in an interval with
a width of roughly 200 ${\rm cm}^{-1}$ . 

Only one vibrational level ($\nu_{\max}=1$) of a single mode is treated
explicitly in the molecular Hamiltonian. This is motivated mainly
by the small value of the Huang-Rhys factor. We have verified that
we obtain similar results including a larger number of vibrational
quanta. The excited-state part of the total Hamiltonian, Eq. \ref{eq:Hamiltonian},
reads

\begin{widetext}

\begin{equation}
\hat{H}_{m}(\Delta E,\, J)=\begin{pmatrix}E & 0 & V_{00} & V_{01}\\
0 & E+\hbar\omega_{0} & V_{01} & V_{11}\\
V_{00} & V_{01} & E+\Delta E & 0\\
V_{01} & V_{11} & 0 & E+\Delta E+\hbar\omega_{0}
\end{pmatrix}\label{eq:H_electronic}
\end{equation}
\end{widetext}where $E$ is the energy of the optical transition
($E$ = 12210 cm$^{-1}$), $V_{\nu\nu'}=J\,(FC)_{\nu}(FC)_{\nu'}$,
$J$ is the coupling between the two BChls and $(FC)_{\nu}=\left<g_{0}\middle|e_{\nu}\right>$
is the Franck-Condon factor characterizing the overlap of the vibrational
wavefunction in the electronic ground state with that of the $\nu^{th}$
vibrational level in the electronic excited state. 

In this work, we study the non-rephasing pathways ($R_{1}$ and $R_{4}$),
and more specifically, their contributions to the signal on the diagonal
of the 2D spectrum. In particular, we focus on $R_{1}^{(1\alpha,g_{0})}$
and $R_{4}^{(\alpha\alpha,g_{1})}$ with $\alpha=2$ for $\omega_{0}\leq135\,{\rm cm^{-1}}$
and $\alpha=3$ otherwise (states numbered with increasing energy),
because, as will be shown below, they both involve the zero- and one-phonon
levels local to molecule 1, either on its electronic excited- or ground-state
manifold. Comparison of the results for these two particular pathways
will enable us to discuss the different characteristic of the vibronic
and the ground-state vibrational coherences, thus taking into account
both recent theories of the origin of the long-lived oscillations
in the 2D spectra of photosynthetic systems \cite{Christensson2012a,Tiwari2012a}. 

Unlike in Ref. \cite{Tiwari2012a}, we concentrate on the oscillations
of the diagonal peaks in the 2D spectrum. This allows us to stay in
one-particle approximation \cite{Philpott1971a}, because, for the
ground-state contributions, we are looking at pathway $R_{4}^{(\alpha\alpha,g_{\nu})}$
(see Fig. \ref{fig:Liouville-pathways}B). Since the state $|\alpha\rangle$
is initially excited on the monomer 1 and originates from the ground
state of the dimer, it cannot contain an excitation of ground-state
vibration on the other monomer. Unlike for the crosspeak, the one-particle
approximation holds on the diagonal cut for all states $|\alpha\rangle$
accessible from the dimer ground state.

\subsection{Influence of coupling and energy gap on the dipole pre-factor }

\begin{figure}
\includegraphics[width=1\columnwidth]{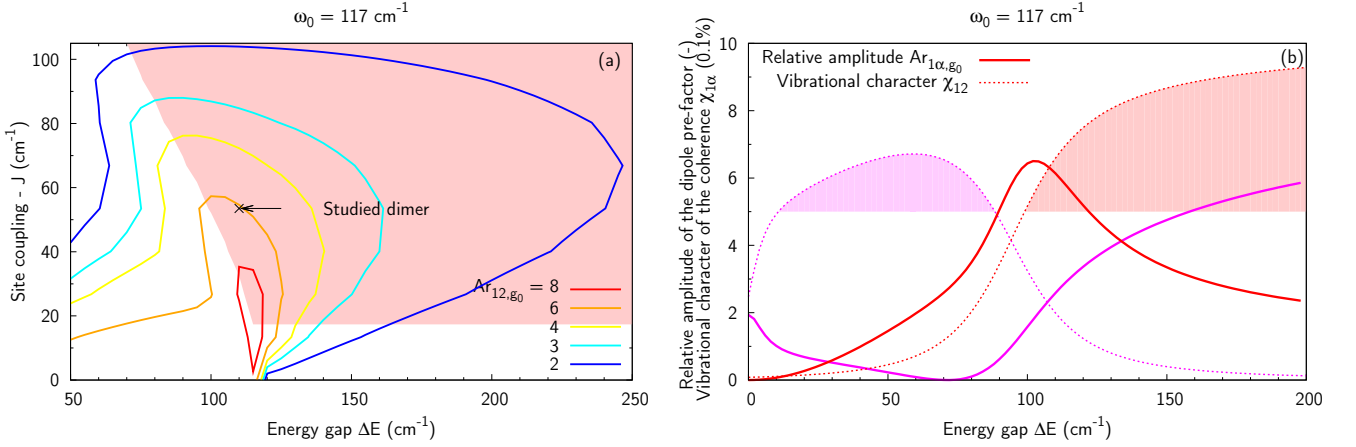}

\caption{Relative amplitude of the dimer dipole moment pre-factor over that
of the monomer $(Ar_{1\alpha,g_{0}})$ for the non-rephasing pathway
$R_{1}$ at initial time (a) involving coherence (1,2) as a function
of the energy gap $\Delta E$ and coupling $J$, (b) involving coherence
(1,$\alpha$), with $\alpha=2$ (red) and $\alpha=3$ (magenta), as
a function of the energy gap $\Delta E$ for $\omega_{0}=117$~cm$^{-1}$.
The plain thick lines represent the amplitudes, the dashed lines denote
the associated measure of the vibronic character $(0.1\,\chi_{1\alpha}$).
In both sub-figures, the prevailingly vibrational character of the
eigenstates ($\chi_{1\alpha}>50\%$) is highlighted with colored areas.
A significant enhancement of the dipole moment pre-factor can be observed
for coherence involving eigenstates 1 and 2, which is strongly vibrational
for $\Delta E>99$ cm$^{-1}$. \label{fig:amplitude}}
\end{figure}

As detailed in Section \ref{sec:2D-Non-Linear-Spectroscopy}, the
initial amplitude of the signal is directly proportional to the dipole
pre-factor (see Eq. \ref{eq:amplitude_R1}). Here, we study the influence
of the energy gap $\Delta E$ and site coupling $J$ on the dipole
pre-factor, with detailed results for pathways $R_{1}$ involving
the zero-phonon level of the excited state on molecule 1 (i.e. eigenstate
1, mainly localized on $|e_{1}g_{0}\rangle$). Fig. \ref{fig:amplitude}
presents the relative transition dipole moment pre-factor for the
dimer response function $R_{1}^{(1\alpha,g_{0})}$ involving electronically
excited eigenstate $|\alpha\rangle=|2\rangle,|3\rangle$ for $\omega_{0}=117$~cm$^{-1}$,
respective to the pre-factor corresponding to the vibrational oscillations
on a monomer:
\begin{equation}
Ar_{1\alpha,g_{0}}=\left(A_{1\alpha,g_{0}}^{(1)}\right)_{\textrm{dimer}}/\left(A_{12,g_{0}}^{(1)}\right)_{\textrm{monomer}}.
\end{equation}
 Fig. \ref{fig:amplitude}a shows results for the coherence involving
the two lowest excited eigenstates $|1\rangle$ and $|2\rangle$.
Compared to the monomer, the amplitude of the dipole pre-factor in
the heterodimer is clearly enhanced (up to more than 8 times) for
an energy gap comparable to the vibrational energy. It should be noted
that, in the range of very small coupling the excitonic model breaks
down, and the electron-phonon coupling might effectively destroy the
excitonic mixing. For moderate resonance coupling values and increasing
its value, the region for which the amplitude is enhanced through
borrowing of transition dipole moment spreads over a wider range of
energy gaps corresponding to a wider mixing region. Note that the
studied excited-state coherence has a prevailingly vibrational character
only in the shaded area of Fig. \ref{fig:amplitude}a. 

Fig. \ref{fig:amplitude}b presents the amplitude of the dipole pre-factor
in the coupled dimer relative to the monomeric one for coherences
$\rho_{1\alpha}$ as a function of the energy gap $\Delta E$, using
$J_{0}=-53.5$~cm$^{-1}$. We also present the local vibrational
character $\chi$ of the coherences (\textit{cf} Eq. \ref{eq:vibchar}).
We can define a characteristic energy gap $\Delta E_{\chi}=94$ cm$^{-1}$
for which the coherences $\rho{}_{12}$ and $\rho_{13}$ are equally
delocalized, i.e. $\chi_{12}=\chi_{13}$. The highest enhancement
($\sim6.5$ times higher than that in a monomer) is obtained for the
coherence involving the two lowest excitons $\left(1,2\right)$. This
coherence is of prevailingly vibrational character for $\Delta E>99$
cm$^{-1}$. The coherence $\rho_{13}$ exhibits only a low enhancement
($<2$) in the region where is it prevailingly vibrational ($10<\Delta E<88$
cm$^{-1}$) and will therefore make minor contributions to long-lived
coherences in the 2D spectrum.

\subsection{Amplitude of the 2D signal and life time of coherences \label{sub:Amplitude}}

In this section, we study the response functions $R_{1}$ and $R_{4}$
involving the vibrational coherence in the electronic excited-state
and in the electronic ground-state manifolds, respectively. The calculations
have been performed on the FMO inspired dimer at 77 K for various
vibrational frequencies, using an overdamped Brownian oscillator mode
bath (see e.g. \cite{MukamelBook}) with the Debye frequency $\Lambda=130$
cm$^{-1}$ and reorganization energy $\lambda=35{\rm \, cm}^{-1}$.
Pigment energies were sampled from a Gaussian distribution with 80
cm$^{-1}$ FWHM centered at the reference energy gap ($\Delta E_{0}=110$
cm$^{-1}$). The energy transfer rates that contribute to the dephasing
of the coherences were calculated from the same energy gap correlation
functions as the line-shape functions (see Appendix) using the standard
Redfield theory. The theory is identical with the one used in Ref.
\cite{Christensson2012a}.

\begin{figure}
\includegraphics[width=1\textwidth]{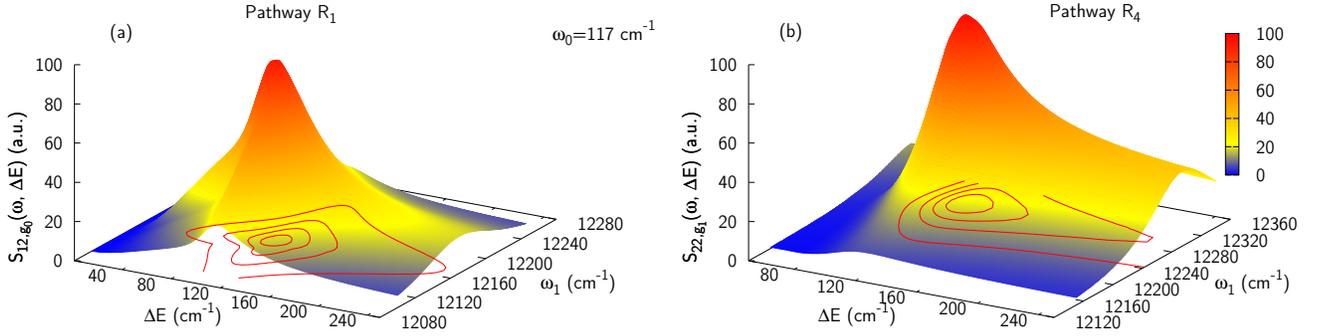}

\caption{Absolute value of the oscillating signal amplitudes $S_{12,g_{0}}^{(1)}$
and $S_{22,g_{1}}^{(4)}$(Eqs. (\ref{eq:S12_1}) and (\ref{eq:S22_4}))
for pathways $R_{1}$ (a) and $R_{4}$ (b) as a function of the energy
gap $\Delta E$ and excitation frequency $\omega_{1}$. In both pathways,
the amplitude is significantly enhanced in the vicinity of the vibrational
mode energy, i.e. for $\Delta E\sim\omega_{0}$ ($\omega_{0}=117{\rm \; cm}^{-1}$
here) due to the resonance occurring. \label{fig:sigdE}}
\end{figure}

In Fig. \ref{fig:sigdE}, we present the energy gap dependence of
the diagonal cut ($\omega_{1}=\omega_{3})$ through the Fourier transform
of the response functions $R_{1}$ and $R_{4}$ before accounting
for the energy disorder \begin{widetext} 
\begin{eqnarray}
S_{12,g_{0}}^{(1)}(\omega_{1},\Delta E) & = & A_{12,g_{0}}^{(1)}(\Delta E)\:\tilde{G}_{1g_{0}}(\omega_{1},\Delta E)\,\max\left(G'_{12}(t_{2},\Delta E)\right)\,\tilde{G}_{1g_{0}}(\omega_{1},\Delta E),\label{eq:S12_1}\\
S_{22,g_{1}}^{(4)}(\omega_{1},\Delta E) & = & A_{22,g_{1}}^{(4)}(\Delta E)\:\tilde{G}_{2g_{0}}(\omega_{1},\Delta E)\,\max\left(G'_{g_{1}g_{0}}(t_{2},\Delta E)\right)\,\tilde{G}_{2g_{0}}(\omega_{1},\Delta E).\label{eq:S22_4}
\end{eqnarray}
\end{widetext} The energy of eigenstates $1$ and $2$ depends on
both $\Delta E$ and $\omega_{0}$ ($\omega_{0}$ is fixed at $117\,{\rm cm^{-1}}$
here). There is a clear resonance occurring when the energy gap $\Delta E$
is comparable to the frequency of the nuclear mode, $\Delta E\sim\omega_{0}$.
This resonance demonstrates the borrowing of dipole moment from the
electronic to the vibrational transition, as suggested by \cite{Christensson2012a}
and illustrated in Section \ref{sec:Enhancement-of-Coherence}. The
oscillating signal amplitude also depends on the line-shape function
$\tilde{G}(\omega)$ which now also depends on the mixing and therefore
on $\Delta E$. This dependence leads to an even larger enhancement
of the oscillation amplitude (with respect to the one observed for
the vibrational oscillations on a monomer) than the transition dipole
moment pre-factor alone would predict. 

\begin{figure}
\includegraphics[width=1\columnwidth]{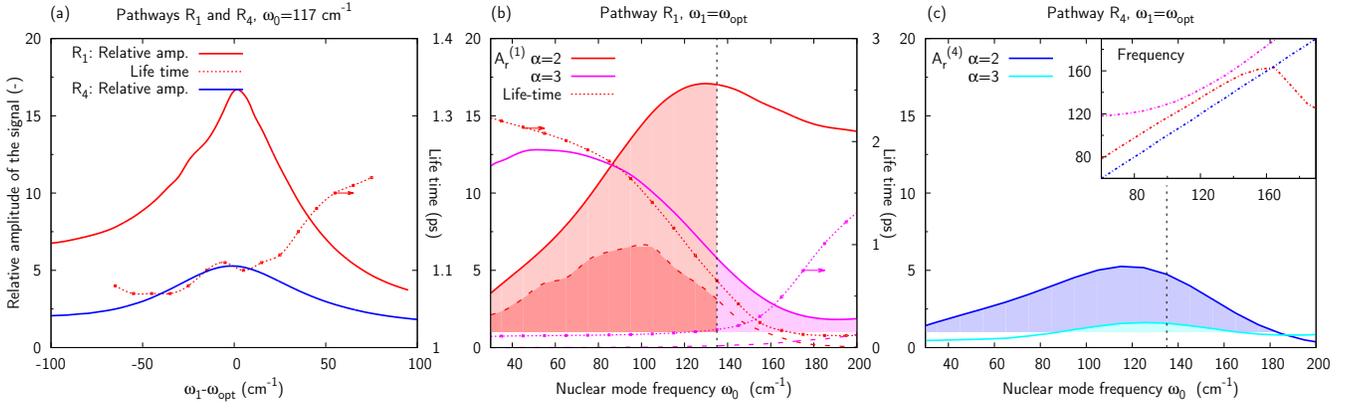}

\caption{(a) Relative amplitude of the signal $A_{r}$ at initial time for
pathways $R_{1}$ and $R_{4}$ (Eqs. (\ref{eq:Ar1}) and (\ref{eq:Ar4}))
involving the prevailingly vibrational coherence $\rho_{12}$ as a
function of the excitation frequency $\omega_{1}$ for $\omega_{0}=117$
cm$^{-1}$. The enhancement of the vibronic coherence (pathway $R_{1}$)
is up to 3 times larger than that of the ground-state vibrational
coherence (pathway $R_{4}$). (b, c) Relative amplitude of the diagonal
signal through pathways $R_{1}$ (b) and $R_{4}$ (c) at $\omega_{1}=\omega_{{\rm opt}}$
as function of the vibrational mode frequency $\omega_{0}$ for $\alpha=2$
(red, blue) and $\alpha=3$ (magenta, cyan). The shaded areas represent
the domains in which the coherence is enhanced and of prevailingly
vibrational character. The dotted dashed lines show the vibronic coherence
life time, and the dashed lines represent the signal relative amplitude
after 1 ps. The inset in (c) presents the frequency of the oscillating
signal through the different pathways. The vertical black dashed line
at $\omega_{0}=135\,{\rm cm^{-1}}$ delimits the mode frequency at
which the ordering of the eigenstates is reorganized and the vibrational
character of the coherences changes. It appears that long-lived ($>$1
ps) vibronic coherences are created for $\omega_{0}<120\,{\rm cm^{-1}}$
(b) whereas any mode generates long-lived ground-state vibrational
coherences (c). \label{fig:AmpTau}}
\end{figure}

The complete theory of enhancement has to include averaging over the
distribution of energy gaps to account for the energy disorder. Fig.
\ref{fig:AmpTau} presents the maximal amplitude of the signal from
pathway $R_{1}$ involving coherence $\rho_{1\alpha}$ relative to
that of the purely vibrational coherence in a monomer ($A_{r}^{(1)}$,
defined below), as well as that from $R_{4}$ ($A_{r}^{(4)}$), 
\begin{eqnarray}
A_{r}^{(1)}(\omega_{1}) & = & \frac{\max[S_{1\alpha,g_{0}}^{(1)}(\omega_{1},t_{2}^{(t')},\omega_{1})]_{{\rm dimer}}}{\max[S_{12,g_{0}}^{(1)}(\omega_{1},t_{2}^{(\max)},\omega_{1})]_{{\rm monomer}}},\label{eq:Ar1}\\
A_{r}^{(4)}(\omega_{1}) & = & \frac{\max[S_{\alpha\alpha,g_{1}}^{(4)}(\omega_{1},t_{2}^{(\max)},\omega_{1})]_{{\rm dimer}}}{\max[S_{22,g_{1}}^{(4)}(\omega_{1},t_{2}^{(\max)},\omega_{1})]_{{\rm monomer}}},\label{eq:Ar4}
\end{eqnarray}
along the diagonal cut of the 2D spectrum ($\omega_{3}=\omega_{1}$).
The letter $\alpha$ denotes the exciton level which is composed mainly
of the first vibrationally excited level located on molecule 1 ($\alpha=2$
for $\omega_{0}\leq135\,{\rm cm^{-1}}$ and 3 otherwise). $t_{2}^{(t')}$
represents the time of the first maximum after a time $t'$ ($t'=0$
or 1 ps in the following) and $t_{2}^{(max)}$ is the time at which
the signal is maximum for non-decaying coherences. Fig. \ref{fig:AmpTau}a
presents results for $\omega_{0}=117$ cm$^{-1}$ ($\alpha=2$) as
a function of the excitation frequency $\omega_{1}$ relatively to
the frequency $\omega_{{\rm opt}}$, which corresponds to the absorption
maximum (for $\omega_{0}=117\,{\rm cm^{-1}}$, $\omega_{{\rm opt}}=12165$
and 12300 cm$^{-1}$ for $R_{1}$ and $R_{4}$, respectively). For
both pathways, the signal amplitude in the coupled dimer is clearly
enhanced compared to that of the corresponding monomer (maximum 5
times higher for $R_{4}$, and more than 16 times for $R_{1}$). 
Fig. \ref{fig:AmpTau}a also shows the life time of coherence $\rho_{12}$
involved in $R_{1}$, obtained from a fit of the calculated signal
-- the life time of the ground-state coherence (pathway $R_{4}$)
is not displayed because the purely vibrational coherences are assumed
non-decaying as in Refs. \cite{Tiwari2012a,Christensson2012a}. Using
$\omega_{0}=117$ cm$^{-1}$, the observed excited-state coherence
appears to survive for slightly more than 1 ps.

Fig. \ref{fig:AmpTau}b shows the maximal enhancement obtained at
initial time (plain lines) and after 1 ps (dashed lines) for different
vibrational frequencies (on the diagonal cut at $\omega_{1}=\omega_{{\rm opt}}$
in the 2D spectrum) for pathway $R_{1}$. Results for pathway $R_{4}$
are presented in Fig. \ref{fig:AmpTau}c. In order to account for
the dependency of the eigenstate composition on the mode frequency
and the resulting reordering of eigenstates at $\omega_{0}=135\,{\rm cm^{-1}}$,
results are presented for coherences $\rho_{12}$ and $\rho_{13}$
involved in both pathways $R_{1}$ and $R_{4}$. For $R_{1}$, the
relevant domain in which the vibronic coherence is of prevailingly
vibrational character $\left(\chi>0.5\right)$, is highlighted with
shaded areas in Fig. \ref{fig:AmpTau}b. For pathway $R_{4}$, the
ground-state vibrational coherence remains purely vibrational independently
of the origin of the excited state which participates in the pathway.
For both pathways, the enhancement is larger when coherence $\rho_{12}$
is involved. Comparing Figs. \ref{fig:AmpTau}b and \ref{fig:AmpTau}c,
it can be seen that, at initial time, the enhancement for pathway
$R_{1}$ is significantly (up to 3 times) larger than that for $R_{4}$.
Also, for both pathways, there is a broad resonance in the enhancement
in the vicinity of the energy gap value $\Delta E_{0}=110\ {\rm cm^{-1}}$,
which is in accordance with the similarly broad resonance for the
enhancement of the ground-state contribution for a crosspeak reported
in Ref. \cite{Tiwari2012a}. Because the excited-state coherences
involve mixing of electronic and vibrational states, their life time
depends, among others, on the dephasing time of the excited states.
Consequently, long-lived (>1 ps) coherences will only be created for
$\omega_{0}<120$ cm$^{-1}$ for the pathway $R_{1}$. In the case
of the pathway $R_{4}$, any nuclear mode will generate long-lived
coherences because the purely vibrational coherences decay slowly.
Therefore for higher mode frequencies, long-lived coherences should
originate from pathway $R_{4}$ only. We studied in detail the $R_{1}$
pathway involving $\rho_{12}$ using $\omega_{0}=185$ cm$^{-1}$
(not presented here). We concluded that, although at initial time
the signal is of comparable amplitude with $\omega_{0}=117$ cm$^{-1}$,
it is almost absent after 1 ps due to the decay. The inset in Fig.
\ref{fig:AmpTau}c shows the frequency of the oscillating signals.
It is verified that the ground-state vibrational coherences oscillate
at the frequency of the nuclear mode, whereas the frequency of the
vibronic coherences does not exactly matches it due to the excitonic
splitting effect and the resulting dependence on the site coupling
and excited-state energies. 

Table \ref{tab:dimer_comp} presents the contributions of the local
basis excitations to the vibronic states averaged over energy disorder
for $\omega_{0}=117$ cm$^{-1}$. Eigenstate 1 consists of 85\% excitation
of the $\nu=0$ transition on site 1 and 15\% on site 2. Eigenstate
2 corresponds to 64\% of the $\nu=1$ transition on site 1. This composition
explains the results presented above, and confirms that prevailingly
vibrational coherences exhibit a prolonged life time. 

\begin{table}
\caption{Composition of the eigenstates and eigenenergies averaged over the
energy disorder for {\footnotesize $\omega_{0}=117$ cm$^{-1}$}.
Significant contributions are highlighted with bold font. }

{\footnotesize }%
\begin{tabular}{cccccccc}
\hline 
\multicolumn{2}{c}{} & \multicolumn{2}{c}{{\footnotesize $\left<|c_{n_{\nu}m_{v'}}^{a}|^{2}\right>_{\Delta}$}} & {\footnotesize $\alpha$=1} & {\footnotesize $\alpha$=2} & {\footnotesize $\alpha$=3} & {\footnotesize $\alpha$=4}\tabularnewline
\hline 
\multirow{2}{*}{{\footnotesize n=e}} & \multirow{2}{*}{{\footnotesize m=g}} & {\footnotesize $\nu$=0} & \multirow{2}{*}{{\footnotesize $\nu'$=0}} & \textbf{\footnotesize 0.85}{\footnotesize{} } & {\footnotesize 0.08} & {\footnotesize 0.07 } & {\footnotesize 0.00 }\tabularnewline
 &  & {\footnotesize $\nu$=1} &  & {\footnotesize 0.00 } & \textbf{\footnotesize 0.64}{\footnotesize{} } & \textbf{\footnotesize 0.36 } & {\footnotesize 0.00 }\tabularnewline
\multirow{2}{*}{{\footnotesize n=g}} & \multirow{2}{*}{{\footnotesize m=e}} & \multirow{2}{*}{{\footnotesize $\nu$=0}} & {\footnotesize $\nu'$=0} & {\footnotesize 0.15 } & \textbf{\footnotesize 0.29 } & \textbf{\footnotesize 0.57}{\footnotesize{} } & {\footnotesize 0.00 }\tabularnewline
 &  &  & {\footnotesize $\nu'$=1} & {\footnotesize 0.00 } & {\footnotesize 0.00 } & {\footnotesize 0.00 } & \textbf{\footnotesize 0.99}{\footnotesize{} }\tabularnewline
\hline 
\multicolumn{2}{c}{{\footnotesize $\left<|\mu_{\alpha g_{0}}|^{2}\right>_{\Delta}$}} &  &  & {\footnotesize 1.64} & {\footnotesize 0.16} & {\footnotesize 0.16} & {\footnotesize 0.00}\tabularnewline
\multicolumn{2}{c}{{\footnotesize $\left<|\mu_{\alpha g_{1}}|^{2}\right>_{\Delta}$}} &  &  & {\footnotesize 0.05} & {\footnotesize 0.48} & {\footnotesize 0.47} & {\footnotesize 0.82}\tabularnewline
\multicolumn{2}{c}{{\footnotesize $\left<\epsilon_{\alpha}\right>_{\Delta}-\left<\epsilon_{1}\right>_{\Delta}$}} & {\footnotesize (cm$^{-1}$)} &  & {\footnotesize 0.} & {\footnotesize 128.} & {\footnotesize 163.} & {\footnotesize 294.}\tabularnewline
\hline 
\end{tabular}\label{tab:dimer_comp}
\end{table}

\begin{figure}
\begin{centering}
\includegraphics[width=1\columnwidth]{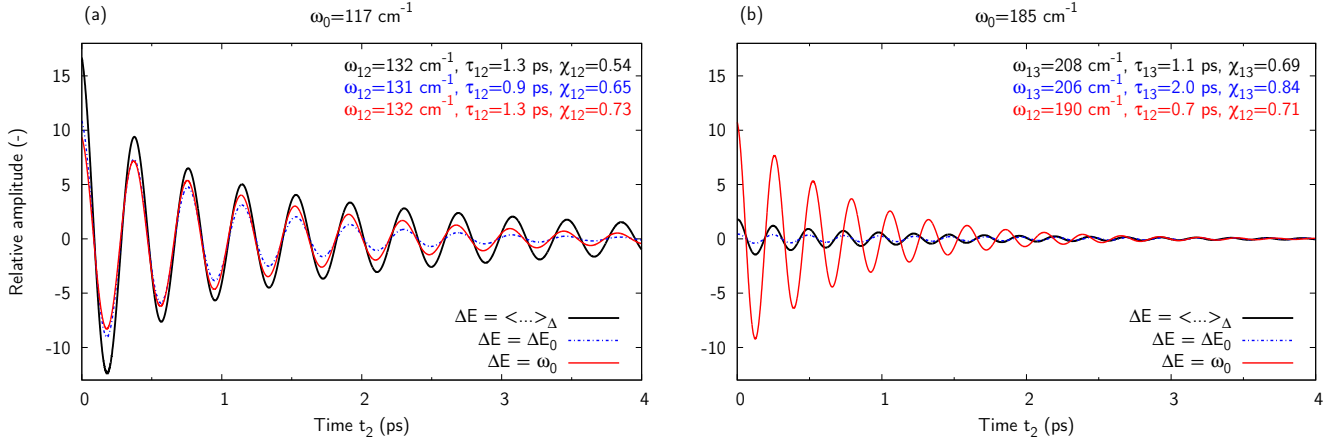} 
\par\end{centering}

\centering{}\caption{Time evolution of the signal involving coherences of prevailingly
vibrational character originating in pathway $R_{1}$ for two mode
frequencies: $\omega_{0}=117$ cm$^{-1}$ (a) and $185$ cm$^{-1}$
(b). The dynamics of the signal has been computed with different energy
gaps, namely the energy gap averaged over the energy disorder (denoted
by $\left<\dots\right>_{\Delta}$), the resonant condition ($\Delta E=\omega_{0}$)
and the reference gap energy ($\Delta E=\Delta E_{0}=110\;{\rm cm}^{-1}$).
The signal amplitude is presented relatively to that of the monomer
in the region where it is maximum ($\omega_{1}=\omega_{{\rm opt}}$).
In part (a) the mode frequency in the vicinity of the energy gap ($\omega_{0}=117$
cm$^{-1}$) creates long-lived oscillations of significant amplitude
even with energy disorder. In part (b), i. e. for a mode frequency
out of resonance with the energy gap ($\omega_{0}=185\,{\rm cm^{-1}}$),
the signal benefits only sightly from the enhancement mechanism, and
it remains of low amplitude. \label{fig:signal}}
\end{figure}

Figs.~\ref{fig:signal}a and \ref{fig:signal}b show the evolution
of the signal (at $\omega_{1}=\omega_{{\rm opt}}$) resulting from
pathway $R_{1}$ and involving the prevailingly vibrational coherence
for two different nuclear mode frequencies ($\omega_{0}=117$ and
185 cm$^{-1}$), with different energy gaps, namely the reference
energy gap ($\Delta E_{0}=110$ cm$^{-1}$), the resonant condition
$\Delta E=\omega_{0}$, and $\Delta E$ sampled from the Gaussian
distribution centered on the reference energy gap (denoted by $\left<\dots\right>_{\Delta}$
in Fig. \ref{fig:signal}). The fitted frequencies and life time of
the signal are indicated in the figures, along with the vibrational
character $\chi$ of the coherence. For the mode frequency $\omega_{0}=117$
cm$^{-1}$, we can expect long-lived oscillations with large amplitudes
for any of the presented energy gaps, because the latter is either
equal to or neighboring the vibrational mode frequency. Fig. \ref{fig:signal}a
confirms this expectation. A different behavior is observed for $\omega_{0}=185$
cm$^{-1}$. Here, the amplitude of the vibronic coherence is significantly
increased (from 0.4 to more than 10 times that in a monomer) when
the energy gap corresponds to the frequency of the nuclear mode, which
illustrates the resonance interaction between the two monomers. However,
because the averaged energy is sampled around $110\,{\rm cm^{-1},}$
the total signal only slightly benefits from the enhancement mechanism
and its amplitude is therefore smaller compared to that obtained using
$\omega_{0}=117\,{\rm cm^{-1}.}$ 

To sum up, the study of a model dimer has enabled us to identify the
resonance between the electronic energy gap and the vibrational frequency
as a crucial condition of the observation of long-lived oscillations
in 2D spectra. The enhancement mechanism lies in intensity borrowing
by excitonic states with strong vibrational character (the vibronic
excitons) from the strongly allowed electronic states. The vibronic
states thus exhibit a significant intensity despite the small Huang-Rhys
factor of their vibrational component. This mechanism leads to an
enhancement of the initial amplitude of both vibronic and ground-state
vibrational coherences over a bandwidth of about 100 cm$^{-1}$. However,
the mixing of the electronic and vibrational character of the vibronic
coherences implies that these coherences experience additional dephasing.
The strong dependence of the life-time enhancement on the mode frequency
means that vibronic coherences will be observable at long population
times only when the frequency of the vibrational mode is close to
the electronic resonance (\textit{cf} Figs. \ref{fig:AmpTau}b and
\ref{fig:signal}). For the present model, this dephasing leads to
similar amplitude of the vibronic and ground-state vibrational coherences
on a picosecond timescale despite the stronger enhancement of the
initial amplitude of the former. It does not therefore seem to be
possible to exclude one or the other type of coherence \emph{a priori
}based on amplitude arguments. Both mechanisms will be most likely
observed simultaneously in an experiment. The most useful difference
between the two types of coherences is the difference in oscillation
frequency (\textit{cf} inset of Fig. \ref{fig:AmpTau}c). The vibrational
coherence oscillates with a frequency which is equal to that of the
vibrational mode, while the frequency of the vibronic-exciton coherence
is shifted due to the excitonic interaction.

\section{Discussion \label{sec:Dimer-vs-FMO}}

The detailed investigation of the interaction between nuclear and
electronic DOF in a model dimer provides more insight into the mechanism
of enhancement of the amplitude and life time of the oscillations
seen in 2D experiments. These results serve as a verification of the
mechanism proposed by Christensson \textit{et al.} \cite{Christensson2012a},
and form a basis for the discussion of this model in relation to the
one proposed by Tiwari et al. \cite{Tiwari2012a}. We point out that
despite technical differences, such as the subtraction of the symmetric
linear combination of the nuclear mode coordinates in Tiwari et al.
\cite{Tiwari2012a}, the Hamiltonian used in these two works is the
same, and so is the basics of the enhancement mechanism. The contribution
of vibrational states to the optical signal is enhanced by their interaction
with electronic DOF. The models differ only in the manifold in which
the oscillations take place. Christensson \textit{et al.} \cite{Christensson2012a}
suggested the excited state, while Tiwari et al. \cite{Tiwari2012a}
studied the ground-state contribution. As we have illustrated above,
the amplitude of both ground-state vibrational and vibronic coherences
is significantly enhanced in a system of coupled molecules as compared
to the isolated monomers. The recent 2D experiments on BChls in solution
did not find any significant vibrational coherences \cite{Fransted2012a}.
These results are in line with what one would expect based on the
Huang-Rhys factors of the low frequency vibrational modes determined
by previous experiments. However, in aggregates, an enhancement by
factor of five at $1$ ps was found for both excited- and ground-state
contributions, strong enough to elevate the signal above the noise
level.

The ability to distinguish ground- and excited-state contributions
to non-linear optical signals is notoriously difficult, and to identify
unique signatures of electronic, vibronic-exciton and vibrational
coherence is less than trivial \cite{Christensson2011a,Turner2012a,Butkus2012a,Mancal2012b}.
One promising proposal has been the use of specific polarization sequences.
It has been shown that a specific combination of the polarizations
in the four wave mixing sequence can single out coherences involving
transitions with non-zero angle in the molecular frame \cite{Zanni2001a,Schlau-Cohen2012a}.
When the vibrational modes are treated as members of the bath (i.e.
standard exciton picture), vibrational transitions on a single pigment
are always parallel, and the orientational average for pathways representing
vibrational coherences will be zero. Therefore such polarization sequences
are termed \textquotedbl{}coherence specific\textquotedbl{} sequences
and have been assumed to only report on electronic coherences \cite{Schlau-Cohen2012a}.
However, when the vibrational modes are treated explicitly, the mixing
between vibrational and electronic DOF leads to a non-zero angle between
the vibronic states, even those largely located on the same pigment.
The vibronic coherences (pathway $R_{1}$) discussed in Ref. \cite{Christensson2012a}
will thus not be eliminated by the coherence specific polarization
sequence. The same argument applies to the ground-state vibrational
coherences. When modes are treated explicitly, there will be a nonzero
angle between $\mu_{\alpha g_{0}}$ and $\mu_{\alpha g_{1}}$, and
pathways of type $R_{4}$ will also contribute to the signal \cite{Tiwari2012a}.
We can thus conclude that the \textquotedbl{}coherence specific\textquotedbl{}
sequence is not able to unambiguously distinguish between electronic,
vibronic and ground-state vibrational coherence.

To be able to determine the origin of the oscillations observed in
the experiments, additional arguments, such as the oscillation frequencies
or dephasing times, are needed. The ground-state vibrational coherences
will oscillate with a frequency equal to that of the relevant vibrational
mode, while for the vibronic coherences, the oscillation frequency
will depend on the vibrational frequency as well as on the electronic
coupling and site energies. The vibronic coherences also experience
additional dephasing due to the mixing of electronic and vibrational
DOF, while the ground-state vibrational coherences only experience
vibrational dephasing. These quantitative differences could in principle
be used to distinguish the two models, if the vibrational frequencies,
Huang-Rhys factors and dephasing times of the vibrational modes were
known.

The low frequency vibrational spectrum of BChl-a in solutions or in
different protein environments have been studied by a number of techniques.
Studies of isolated BChl-a in frozen solution have demonstrated that
the strongest vibrational mode is found around 180-190 cm$^{-1}$
\cite{Ratsep2011a,Zazubovich2001a,Renge1987a,Diers1995a}. These experiments
have identified a mode around 160 cm$^{-1}$, but its Huang-Rhys factor
was found to be significantly weaker than the one of the modes around
180-190 cm$^{-1}$ \cite{Ratsep2011a,Zazubovich2001a}. Although the
results of the experiments are rather consistent, it should be pointed
out that the low-frequency spectral range is sensitive to the local
conformation of the macrocycle and the coordination of the central
Mg atom. This effect was illustrated by R\"{a}tsep \textit{et al.}
\cite{Ratsep2011a} by measuring the FLN spectra of BChl-a in different
solvents. These experiments demonstrate that results obtained on the
isolated pigment might not be a good indicator of the frequencies
and Huang-Rhys factors observed in the protein environment.

A widely studies case of \textquotedbl{}isolated\textquotedbl{} BChl-a
in a protein environment is the accessory BChl in the bacterial reaction
center. Resonance Raman (RR) spectra of the accessory BChl (B-band)
have been reported by several groups \cite{Cherepy1995a,Cherepy1997a,Czarnecki1997a}.
The room temperature spectrum identified modes at 84, 117, 180 and
210 cm$^{-1}$, where the latter two were considerably stronger \cite{Cherepy1997a}.
Measurements at 95 K revealed an additional mode at 160 cm$^{-1}$
which was $\sim$10 times weaker than the mode at 180 cm$^{-1}$ \cite{Cherepy1997a}.
Similar results were obtained at even lower temperature, showing that
modes around 180 and 220 cm$^{-1}$ dominates the spectrum \cite{Czarnecki1997a}.

There is only a limited number of studies on the vibrational spectrum
of the pigments in the FMO complex. A hole burning (HB) study found
two modes at 120 and 160 cm$^{-1}$ with a comparable strength, but
were not able to observe any modes with higher frequencies due to
the narrow absorption band in FMO \cite{Matsuzaki2000a}. Fluorescence
line narrowing (FLN), on the other hand, revealed a spectrum which
was similar to the RR spectrum of the B-band in the reaction center
\cite{Wendling2000a,Ratsep2007a}. The two FLN measurements on the
different species are similar, showing a mode at 117 cm$^{-1}$ and
several modes between 167-202 cm$^{-1}$. The strongest modes in both
cases were found around 190 cm$^{-1}$.

Taking all of the data on the low frequency vibrational spectrum of
BChl-a, we can conclude that none of the reported spectra show the
presence of a single strong mode in the frequency range corresponding
to that observed in the 2D experiments (160 cm$^{-1}$). Rather, the
spectra reveal a large number of vibrational modes with comparable
Huang-Rhys factors in the range from 80-240 cm$^{-1}$ corresponding
to typical energy splittings in FMO \cite{Adolphs2006a}. All these
modes will thus, to a certain extend, experience enhancement due to
the interaction between electronic and vibrational DOF. For the ground-state
wave-packets, the frequency of the oscillation will match that of
the vibrational mode. Based on the discussion above, we would predict
that the mode at 117 cm$^{-1}$ experiences the strongest enhancement.
A rather broad amplitude enhancement curve furthermore implies that
multiple oscillation frequencies should be observed. For the vibronic
coherences, the observed frequency depends on the resonance coupling
as well as on the vibrational frequencies, the number of pigments
and probably also on the number of vibrational modes included in the
Hamiltonian. For the dimer used here, we predict an oscillating frequency
of 132 cm$^{-1}$ and for the full model, the value of 140 cm$^{-1}$
was found \cite{Christensson2012a}. Increasing the resonance coupling
would blue-shift the oscillation frequency further and increase the
amplitude of the vibronic coherence. However, due to the increased
electronic character, these coherences would also experience a stronger
dephasing and a correspondingly shorter life time. Because the amplitude
enhancement, oscillation frequencies and dephasing dynamics are all
closely related to the electronic structure, a simulation including
all pigments is required to determine whether it is possible to find
a set of couplings which results in a strong enhancement, correct
oscillation frequency and a long life time. Such a simulation could
also determine whether vibronic or ground-state vibrational coherences
dominate in the case of FMO. Furthermore, it would be possible to
elucidate why a single oscillation dominates in the 2D spectrum, although
multiple vibrational modes are expected to experience significant
amplitude enhancement according to the reduced dimer models. 

Finally, an enhancement of exciton sizes and energy transport rates
resulting from vibronic coupling has recently been demonstrated in
the interacting pigments of two nearly identical light-harvesting
proteins \cite{Womick2011a}. It was shown that the vibronic-exciton
model provides a realistic explanation for an enhancement in the kinetics
of the system with respect to the purely excitonic model. A robust
enhancement of the transfer rates and exciton sizes was observed at
the resonant condition $\Delta E=\hbar\omega$. The mixing of vibrational
and electronic DOF in molecular aggregates is therefore implied not
only in the enhancement of the observed coherence lifetime, but can
also lead to enhanced rates of energy transport.

\section{Conclusions}

In this work, we have demonstrated on a model dimer how the interaction
between electronic and nuclear degrees of freedom leads to enhanced
amplitudes of vibronic and ground-state vibrational coherences in
2D spectra. The vibronic-exciton model used here also provides, without
contradiction with molecular dynamics simulations, a plausible explanation
for correlated fluctuations in different spectral regions, often postulated
in purely excitonic model to reproduce long-lived coherences. The
enhancement of coherence amplitude requires both sufficient electronic
coupling and a resonance between the vibrational level on one pigment
and the electronic transition on the other. Enhancement is found to
be even more pronounced when the inhomogeneity of the sites is included.
The resonance enhancement mechanism affects both the excited-state
vibronic-exciton coherences as well as the ground-state vibrational
coherences. For the dimer model used in this work, we find that both
types of coherences have a similar magnitude at long (picosecond)
population times. Whether the vibronic coherences dominate in 2D spectra
or not depends on the excited-state dephasing and population dynamics
as well as the influence of inhomogeneous broadening. The results
obtained here highlight the properties of the enhancement mechanism.
To decide which type of coherence dominates in FMO complex would require
simulations with all seven pigments and a full spectral density including
a larger spectrum of vibrational modes. 
\begin{acknowledgments}
This work was supported by the Czech Science Foundation (GACR) grant
nr. 205/10/0989 and the Ministry of Education, Youth and Sports of
the Czech Republic grant MEB 061107, the Austrian Science Foundation
(FWF) within project P22331 and OeAD. The authors would like to thank
Prof. David Jonas for providing us with the manuscript of Ref. \cite{Tiwari2012a}
prior to publication. 
\end{acknowledgments}
\appendix

\section{Third-Order Response Function\label{sec:Third-Order-Response-Function}}

In this Appendix, we detail the dipole pre-factors and time-dependent
response functions for each Liouville pathways $R_{1}-R_{4}$. 

The transition dipole moment pre-factors read 
\begin{eqnarray}
A_{\alpha\beta,g_{\nu}}^{(1)} & = & \langle(\bm{\mu}_{g_{\nu}\alpha}\cdot\bm{e}_{4})(\bm{\mu}_{\beta g_{\nu}}\cdot\bm{e}_{3})(\bm{\mu}_{g_{0}\beta}\cdot\bm{e}_{2})(\bm{\mu}_{\alpha g_{0}}\cdot\bm{e}_{1})\rangle_{\Omega}\label{eq:amplitude_R1-1}\\
A_{\alpha\beta,g_{\nu}}^{(2)} & = & \langle(\bm{\mu}_{g_{\nu}\beta}\cdot\bm{e}_{4})(\bm{\mu}_{\alpha g_{\nu}}\cdot\bm{e}_{3})(\bm{\mu}_{\beta g_{0}}\cdot\bm{e}_{2})(\bm{\mu}_{g_{0}\alpha}\cdot\bm{e}_{1})\rangle_{\Omega}\\
A_{\alpha\beta,g_{\nu}}^{(3)} & = & \langle(\bm{\mu}_{g_{\nu}\beta}\cdot\bm{e}_{4})(\bm{\mu}_{\beta g_{0}}\cdot\bm{e}_{3})(\bm{\mu}_{\alpha g_{\nu}}\cdot\bm{e}_{2})(\bm{\mu}_{g_{0}\alpha}\cdot\bm{e}_{1})\rangle_{\Omega}\\
A_{\alpha\beta,g_{\nu}}^{(4)} & = & \langle(\bm{\mu}_{g_{0}\beta}\cdot\bm{e}_{4})(\bm{\mu}_{\beta g_{\nu}}\cdot\bm{e}_{3})(\bm{\mu}_{g_{\nu}\alpha}\cdot\bm{e}_{2})(\bm{\mu}_{\alpha g_{0}}\cdot\bm{e}_{1})\rangle_{\Omega},
\end{eqnarray}
where $\left<\dots\right>_{\Omega}$ denotes the averaging over a
random orientation of the pigments. 

The time-dependent response functions read:
\begin{eqnarray}
R_{\alpha\beta,g_{\nu}}^{(1)}(t_{3},t_{2},t_{1}) & = & \left<A_{\alpha\beta,g_{\nu}}^{(1)}\, G_{\alpha g_{\nu}}(t_{3})\, G'_{\alpha\beta}(t_{2})\, G_{\alpha g_{0}}(t_{1})\right>_{\Delta}\\
R_{\alpha\beta,g_{\nu}}^{(2)}(t_{3},t_{2},t_{1}) & = & \left<A_{\alpha\beta,g_{\nu}}^{(2)}\, G_{\beta g_{\nu}}(t_{3})\, G'_{\beta\alpha}(t_{2})\, G_{g_{0}\alpha}(t_{1})\right>_{\Delta}\\
R_{\alpha\beta,g_{\nu}}^{(3)}(t_{3},t_{2},t_{1}) & = & \left<A_{\alpha\beta,g_{\nu}}^{(3)}\, G_{\beta g_{\nu}}(t_{3})\, G'_{g_{0}g_{\nu}}(t_{2})\, G_{g_{0}\alpha}(t_{1})\right>_{\Delta}\\
R_{\alpha\beta,g_{\nu}}^{(4)}(t_{3},t_{2},t_{1}) & = & \left<A_{\alpha\beta,g_{\nu}}^{(4)}\, G_{\beta g_{0}}(t_{3})\, G'_{g_{\nu}g_{0}}(t_{2})\, G_{\alpha g_{0}}(t_{1})\right>_{\Delta},
\end{eqnarray}
with $\left<\dots\right>_{\Delta}$ representing the averaging over
a random distribution of pigment energies. The evolution propagators
of the optical and excited-state coherences, $G(t)$ and $G'(t)$,
respectively, read
\begin{equation}
G_{\alpha g_{\nu}}(t)=e^{-i\omega_{\alpha g_{\nu}}t}e^{-\Gamma_{\alpha}t}e^{-\gamma_{\alpha\alpha}g(t)},
\end{equation}
and
\begin{equation}
G'_{\alpha\beta}(t)=e^{-i\omega_{\alpha\beta}\, t}e^{-(\Gamma_{\alpha}+\Gamma_{\beta})\, t}e^{-(\gamma_{\alpha\alpha}+\gamma_{\beta\beta}-2\gamma_{\alpha\beta})\, g(t)},
\end{equation}
where $\omega_{\alpha\beta}=\frac{\epsilon_{\alpha}-\epsilon_{\beta}}{\hbar}$
is the energy difference between the two eigenstates $\alpha$ and
$\beta$, $\Gamma_{\alpha}$ is the eigenstate relaxation rate computed
from the Redfield tensor and $g_{\alpha\beta}(t)=\gamma_{\alpha\beta}\; g(t)$
denotes the line-shape function. The $\gamma$ coefficients can be
expressed via the expansion coefficients $c_{n_{\nu}m_{\nu'}}^{\alpha}$
as

\begin{equation}
\gamma_{\alpha\beta}=\sum_{n,\nu}\left(c_{n_{\nu}m_{0}}^{\alpha}\right)^{2}\left(c_{n_{\nu}m_{0}}^{\beta}\right)^{2}+\sum_{n,\nu>\nu'}\left[\left(c_{n_{\nu}m_{0}}^{\alpha}\right)^{2}\left(c_{n_{\nu'}m_{0}}^{\beta}\right)^{2}+\left(c_{n_{\nu'}m_{0}}^{\alpha}\right)^{2}\left(c_{n_{\nu}m_{0}}^{\beta}\right)^{2}\right].
\end{equation}
Using the Debye spectral density, $\tilde{C}''(\omega)=\pi\omega^{2}J(\omega)=2\lambda\frac{\omega\Lambda}{\omega^{2}+\Lambda^{2}}$,
where $\lambda$ is the bath reorganization energy and $\Lambda$
the Debye frequency, the line-shape function reads, in high-temperature
approximation, \cite{MukamelBook} 

\begin{equation}
g(t)=\left(\frac{2\lambda k_{B}T}{\hbar\Lambda}-i\frac{\lambda}{\Lambda}\right)\left(e^{-\Lambda t}+\Lambda t-1\right),
\end{equation}
with $k_{B}$ the Boltzmann constant and $T$ the temperature.

The contributions of a particular coherence to the signal in a 2D
spectrum is obtained after Fourier transformation of the evolution
propagators during the time intervals $t_{1}$ and $t_{3}$:

\begin{eqnarray}
S_{\alpha\beta,g_{\nu}}^{(1)}(\omega_{3},t_{2},\omega_{1}) & = & \left<A_{\alpha\beta,g_{\nu}}^{(1)}\:\tilde{G}_{\alpha g_{\nu}}(\omega_{3})\, G'_{\alpha\beta}(t_{2})\,\tilde{G}_{\alpha g_{0}}(\omega_{1})\right>_{\Delta}\\
S_{\alpha\beta,g_{\nu}}^{(2)}(\omega_{3},t_{2},\omega_{1}) & = & \left<A_{\alpha\beta,g_{\nu}}^{(2)}\,\tilde{G}_{\beta g_{\nu}}(\omega_{3})\, G'_{\beta\alpha}(t_{2})\,\tilde{G}_{g_{0}\alpha}(\omega_{1})\right>_{\Delta}\\
S_{\alpha\beta,g_{\nu}}^{(3)}(\omega_{3},t_{2},\omega_{1}) & = & \left<A_{\alpha\beta,g_{\nu}}^{(3)}\,\tilde{G}_{\beta g_{\nu}}(\omega_{3})\, G'_{g_{0}g_{\nu}}(t_{2})\,\tilde{G}_{g_{0}\alpha}(\omega_{1})\right>_{\Delta}\\
S_{\alpha\beta,g_{\nu}}^{(4)}(\omega_{3},t_{2},\omega_{1}) & = & \left<A_{\alpha\beta,g_{\nu}}^{(4)}\,\tilde{G}_{\beta g_{0}}(\omega_{3})\, G'_{g_{\nu}g_{0}}(t_{2})\,\tilde{G}_{\alpha g_{0}}(\omega_{1})\right>_{\Delta}
\end{eqnarray}
where $\tilde{G}_{\alpha g_{\nu}}(\omega)=\int_{0}^{+\infty}dt\, e^{i(\omega-\omega_{\alpha g_{\nu}})t-\Gamma_{\alpha}t-\gamma_{\alpha\alpha}g(t)}$
and $\omega_{1},$ $\omega_{3}$ are the excitation and probing frequencies,
respectively. After analytical integration, we obtain 

\begin{equation}
\tilde{G}_{\alpha g_{\nu}}(\omega)=\frac{e^{A_{0}}}{\Lambda A_{0}^{A_{\alpha}\left(\Delta\omega\right)}}\left[\Gamma\left(A_{\alpha}\left(\Delta\omega\right)\right)-\Gamma\left(A_{\alpha}\left(\Delta\omega\right),\, A_{0}\right)\right],
\end{equation}
where $\Delta\omega=\omega_{\alpha g_{\nu}}-\omega$ and $A_{0}=\frac{2\lambda k_{B}T}{\Lambda^{2}}-i\frac{\lambda}{\Lambda}$
are constants, and $A_{\alpha}(\omega)=\frac{\Gamma_{\alpha}-i\omega}{\Lambda}-A_{0}$.
$\Gamma(z)$ is the Euler gamma function with complex argument which
satisfies $\Gamma(z)=\int_{0}^{+\infty}t^{z-1}e^{-t}dt$ and $\Gamma(\alpha,z)=\int_{z}^{+\infty}t^{a-1}e^{-t}dt$
is the incomplete gamma function.

\bibliographystyle{prsty}

\end{document}